\documentclass[11pt,a4paper]{article}
\usepackage{latexsym}
\newlength{\extraspace}
\newlength{\extraspaces}
\newcommand{\be}{\begin{equation}
\addtolength{\abovedisplayskip}{\extraspaces}
\addtolength{\belowdisplayskip}{\extraspaces}
\addtolength{\abovedisplayshortskip}{\extraspace}
\addtolength{\belowdisplayshortskip}{\extraspace}}
\newcommand{\ee}{\end{equation}}

\newcommand{\bea}{\begin{eqnarray}
\addtolength{\abovedisplayskip}{\extraspaces}
\addtolength{\belowdisplayskip}{\extraspaces}
\addtolength{\abovedisplayshortskip}{\extraspace}
\addtolength{\belowdisplayshortskip}{\extraspace}}
\newcommand{\eea}{\end{eqnarray}}

\newcommand{\newsection}[1]{
\vspace{15mm}
\pagebreak[3]
\addtocounter{section}{1}
\setcounter{equation}{0}
\setcounter{subsection}{0}
\setcounter{footnote}{0}
\begin{flushleft}
{\large\bf \thesection. #1}
\end{flushleft}
\nopagebreak
\medskip
\nopagebreak}

\newcommand{\nn}{\nonumber \\[1.5mm]}

\newcommand{\uf}\uparrow
\newcommand{\df}\downarrow
\newcommand{\Td}\bigtriangledown

\def\({\left(}
\def\){\right)}
\def\[{\left[}
\def\]{\right]}
\def\r.{\right.}
\def\l.{\left.}
\def\o{\over}
\def\a{\alpha}

\def\d{\delta}

\def\tp{\tau^{ph}}
\def\D{\Delta}
\def\C{\Gamma}

\def\U{\Omega_{\mu}}

\def\w{\wedge}
\def\dj{{\d\o\d j}}
\def\dJ{{\d\o\d J}}
\def\dV{{\d\o\d V}}
\def\dW{{\d\o\d W}}
\def\dF{{\d\o\d \Phi}}
\def\df{{\d\o\d \fp}}
\def\db{{\d\o\d b}}
\def\j{j^{ph}}
\def\Z{Z_c}
\def\hZ{\hat\Z}
\def\bZ{\bar\Z}
\def\fp{f^{ph}}
\def\co{\Gamma_{gf}}

\def\cq{\bar{\C}}
\def\dX{{\d\o\d\Xi}}

\def\S{{\cal S}}
\def\nv{\nabla_{ V_{\mu}}}
\def\na{\nabla_{A_{\mu}}}

\begin{document}
%\Title{\vtop{\hbox{hep-th/9907092}
\begin{flushright}
\begin{tabular}{l}
GEF-TH-04/994 
\end{tabular}
\end{flushright}

\vskip 1truecm
\centerline{\large\bf Further comments on the background field method}
\bigskip
\centerline{\large\bf and gauge invariant effective actions.}
\bigskip
\centerline{Carlo Becchi\footnote{E-mail: becchi@ge.infn.it}}
\vskip 4pt
\centerline{Renzo Collina\footnote{E-mail: collina@ge.infn.it}}
\vskip 4pt

\centerline{\it Dipartimento di Fisica dell' Universit\'a di Genova}
\centerline{\it and  INFN, Sezione di Genova,}
\centerline{\it Via Dodecaneso 33, I-16146, Genova, Italy}
\vskip 7pt

%\vskip 7pt
 \medskip
\centerline{ABSTRACT}

The aim of this paper is to give a firm and clear proof of the
existence in the background field framework of a gauge invariant
effective action for any gauge theory ({\it background gauge
equivalence}).  Here by effective action we mean a functional whose
Legendre transform restricted to the physical shell generates the
matrix elements of the connected $S$-matrix.  We resume and clarify a
former argument due to Abbott, Grisaru and Schaefer based on the
gauge-artifact nature of the background fields and on the
identification of the gauge invariant effective action with the
generator of the proper, background field, vertices.  \vfill \eject

\newsection{Introduction}

The  analysis at LEP of the effective couplings of the intermediate
bosons of the electro-weak interactions has further increased the need
of an efficient parametrization method for the low energy effective
actions of gauge theories.

In general these effective actions are identified with the functionals
generating the fully renormalized vertices and propagators
contributing to the skeleton graphs, technically speaking, the proper,
1-particle irreducible, amplitudes. Therefore they are controlled by the
non-linear Slavnov-Taylor identity accounting for the BRS symmetry of
the gauge theories.

Since the introduction by De Witt \cite{dW} of the background field
method it is believed that, computing the $S$ matrix, the above
mentioned effective actions are equivalent to those generating the
background field
amplitudes, that is the amplitudes with only background field external
legs.  The background field effective actions are gauge invariant and
hence allow a much simpler parametrization of couplings than the
BRS invariant quantum field effective actions.

The first proofs of this {\it background gauge equivalence}, due to
De Witt and `t Hooft \cite{tH}, were limited to the first loop approximation;
however more recently Abbott \cite{A} has introduced a complete and consistent
renormalization method, based on the background field gauge fixing, and
implementing the gauge invariance under background gauge
transformations to all perturbative orders.  In this framework Abbott,
Grisaru and Schaefer (AGS \cite{AGS}) have suggested how the proof of the
background gauge equivalence could be extended to all orders of
perturbation theory; however in our opinion a definite and clear
proof of this equivalence is still laking.  The purpose of this paper
if to fill this gap.

The argument used by AGS in their proof is that the background field
is introduced as a gauge-artifact and hence the $S$-matrix should not
depend on its choice. They refer to a pure Yang-Mills theory for
which the definition of the $S$-matrix is out of reach, since the
scattering amplitudes are affected with perturbative and
non-perturbative infra-red singularities. However the argument is
general enough to be directly extended to any gauge theory.

To make our proof as clear and firm as possible, we shall refer to an $SU(2)$
Higgs model, for which the $S$-matrix is defined  in perturbation
theory, specifying a complete set of renormalization conditions. We
shall also make extended use of the functional formalism that, as it
is now well known, allows a precise translation of  the diagrammatic
framework in which the original AGS argument was formulated.

First of all we give a precise idea of the AGS argument and of
the open points in the original proof.

For this we have to recall some general fact
about the functional method \cite{BIIM}. Let $j$ label a system of sources
of the
 quantized fields $\phi$, and
$\tau$ label the external fields coupled to a system of composite operators.
We define by  $\Z\[j,\tau\]$ the functional generator of the
connected amplitudes,  those corresponding to connected Feynman
graphs. Under the condition: $\dj\Z |_{j=\tau =0}=0$ , the Legendre
transform $\C$ of $\Z\[j,\tau\]$ is given by :
\be
\Z\[j,\tau \] =-\C\[{\d\o\d j}\Z ,\tau\]+\int j\dj\Z\ .\label{lt1}
\ee

$\C$ is the functional generator of the one-particle-irreducible
amplitudes and identifies a natural choice for an effective action of the
theory; in fact (\ref{lt1}) means that a generic connected
amplitude can be written in the form of a tree graph whose lines and
vertices are defined from the functional derivatives of $\C$.
Assuming the second $j$-derivative of $\Z$, that is the full
propagator, not to be degenerate, one has

\be j=\df\C\[\dj\Z,\tau\]\ ,\label{lt2}\ee

\be {\d\o\d\tau}\Z [j, \tau]=-{\d\o\d\tau}\C \[\dj\Z,
\tau\]\ .\label{ee}\ee
It is worth noticing here that, solving (\ref{lt2})  with the initial
condition $\Z [0,0]=0$, one gets  $\Z\[j,\tau\]$ satisfying (\ref{lt1}),
since both (\ref{lt2}) and (\ref{lt1}) identify uniquely the connected
 functional.

Whenever the theory is infra-red safe, one can introduce the asymptotic
field sources $j^{as}$
 and use  LSZ reduction formulae \cite{BD}.
This can be achieved by a translation of the field sources $j$. For
example, in the case of a scalar field with mass $m$ one has: \be
j(x)\rightarrow j(x) + j^{as}(x)\equiv j(x) +{\cal Z}^{-1}\int d^4yd^4z\
K(x-z; m) \D_+(z-y;
m)\ f^{as}\ (y)\ .\label{Jas}
\ee
where ${\cal Z}$ is a normalization constant and  $\Delta_+(y-z; m)$
is the positive frequency part of the
 Wightman function of the free scalar field\footnote{given by $$
\Delta_+(y-z; m) = \int {d^3k\o (2\pi)^3}{e^{-ik(y-z)}\o 2\omega_k}
= \int {d^4k\o (2\pi)^3}\d\(k^2 - m^2\)\theta \(k^0\)e^{-ik(y-z)}
$$}
and $K(z-x; m) \equiv \(\Box + m^2\)\d^4(z-x) $ .
In general the same formula holds replacing ${\cal Z}$ , $\Delta_+$ and
$K$ with
matrices in the field components, and (\ref{Jas}) can be written in the
form:
\be j^{as}\equiv {\cal Z}^{-1} K*\D_{+}* f^{as}\ .\label{jas}\ee
The introduction of the asymptotic sources allows to define the
elements of the connected scattering matrix $S_c$ according to:
\be
S_c = \Z\[j^{as};\tau\]\ .
\ee
In general only a subset of the asymptotic fields correspond  to physical
particles
and only some of  the composite operators have physical
meaning. Thus one has to select the physical matrix elements of $S_{c}$,
that is:
\be
S^{ph}_c \equiv \Z\[ j^{ph}; \tp \]\ .
\label{MS}
\ee On account of (\ref{lt1}) one has also:\be
S^{ph}_c = -\C\[\dj\Z\[ j^{ph};\tp \], \tp \] +\int j^{ph}\dj\Z\[
j^{ph};\tp \]\ .
\label{MS2}\ee

In the framework of  gauge theories, introducing the background field
$V_\mu$ according to Abbott's all-order scheme, $V_{\mu}$ is identified  with
the source of a composite operator (such as $\tau$) that appears in the
gauge fixing.  If $j_{\mu}$ is the source of the gauge field $A_{\mu}$
and we {\it assume} the existence of the corresponding asymptotic
field and source $j_{\mu}^{as}$ whose restriction to physics is
$j_{\mu}^{ph}$, we can define $S^{ph}_c $ according to (\ref{MS}) and we
get:
\be
{\d \o \d V_{\mu}}\Z\[j_{\mu}^{ph}; V_{\mu}, \tp  \]=0 \ ,\label{ind}
\ee since the background field is
a gauge fixing parameter.
In this framework (\ref{lt1}) corresponds to:
\be
\Z\[j_{\mu}; V_{\mu}, \tau \] = -\C\[{\d \o \d j_{\mu}}\Z\[j_{\mu};
V_{\mu}, \tau \];
V_{\mu}, \tau  \] + \int j_{\mu}{\d \o \d j_{\mu}}\Z\[j_{\mu};
V_{\mu}, \tau \]\ ,
\label{D1}
\ee and hence, using (\ref{MS2}) we can write:
\be
S^{ph}_c = - \C\[{\d \o \d j_{\mu}}\Z\[j_{\mu}^{ph}; V_{\mu}, \tp \] ;
V_{\mu} , \tp \]+\int j_{\mu}^{ph}{\d \o \d j_{\mu}}\Z\[j_{\mu}^{ph};
V_{\mu}, \tp \] \ .
\label{MS0}
\ee
Now,  from (\ref{ee}) and (\ref{ind}) we have:
\be
{\d \o \d V_{\mu}}\Z\[j_{\mu}^{ph}; V_{\mu}, \tp \]=-{\d\o \d
V_{\mu}}\C\[{\d \o \d j_{\mu}}\Z\[j_{\mu}^{ph}; V_{\mu} , \tp\]; V_{\mu} ,
\tp\]=0 \ ,\label{ind1}\ee
meaning that in (\ref{MS0}) we can, more or less, arbitrarily change
the variable $V_\mu$ in $\C$ (however not in $\Z$). Therefore
replacing:
$V_\mu\rightarrow {\d\Z\o \d J_\mu}$, we can write:
\be
S_c = - \C\[{\d \o \d j_{\mu}}\Z\[j_{\mu}^{ph}; V_{\mu} , \tp \]; {\d \o \d
j_{\mu}}\Z\[j_{\mu}^{ph}; V_{\mu} , \tp\] , \tp \] +\int j_{\mu}^{ph}{\d \o \d
j_{\mu}}
\Z\[j_{\mu}^{ph}; V_{\mu}, \tp \] \ .
\label{MS1}
\ee
Formally this equation can be interpreted as equivalent to (\ref{MS0})
after the substitution:
\be \C\[A_\mu ; V_\mu \]\rightarrow \C\[A_\mu ; A_\mu \]\ .\label{sub}\ee
On account of the gauge invariance of $\C\[A_\mu ; A_\mu \]$
\cite{AGS} this could be
interpreted,
following AGS, as a general
 background gauge equivalence proof. More precisely, if $\Z$ in
(\ref{MS1}) were
solution of the equation:
\bea
&&\Z\[j_{\mu}\] = -\C\[{\d \o \d j_{\mu}}\Z\[j_{\mu}\]; {\d \o \d
j_{\mu}}\Z\[j_{\mu}\] \]\nn
&&   + \int
j_{\mu}{\d \o \d j_{\mu}}\Z\[j_{\mu}\]\ , \label{SV0}
\eea
(\ref{MS1})  would give the proof of the existence of a gauge invariant
effective action (in fact $ \C\[A_\mu ; A_\mu \]$) for our gauge theory.
This is however not the case; in particular $ \C\[A_\mu ; A_\mu \]$ cannot
contain any gauge fixing term;  the term existing in
$
\C\[A_\mu ; V_\mu \]$
has been cancelled by the substitution (\ref{sub}):($\C_{G.F.}\[A_{\mu};
A_{\mu} \]
=\left.{\a\o
2}\left(\nv(A_{\mu}-V_{\mu})\right)^2\right|_{V_{\mu} = A_{\mu}}\equiv 0$).
 Therefore (\ref{SV0})
is singular
and the former interpretation of
(\ref{MS1})  difficult to verify.

The natural way to overcome this difficulty would be to start from the
effective action:
\be \C\[A_{\mu};
A_{\mu}, \tau\]+\C_{G.F.}\[A_{\mu}; 0\]\ ,\label{ea}\ee
 and to define the corresponding connected functional generator according to:
\bea
&&\bZ\[j_{\mu}; \tau\] = -\C\[{\d \o \d j_{\mu}}\bZ\[j_{\mu};
\tau\]; {\d \o \d
j_{\mu}}\bZ\[j_{\mu}; \tau\], \tau\]\nn
&&-\C_{G.F.}\[{\d \o \d j_{\mu}}\bZ\[j_{\mu}; \tau\]; 0\]   + \int
j_{\mu}{\d \o \d j_{\mu}}\bZ\[j_{\mu};
\tau\]\ , \label{SV}
\eea
proving the identity:
\be
\bZ\[ j_{\mu}^{ph};\tp\]\equiv \Z\[j_{\mu}^{ph}; 0 , \tp \]
\label{AGS}
\ee
between the solution of (\ref{SV})  and that of (\ref{D1}). We
shall follow this line in next sections.
In particular in section 2 we shall describe the SU(2)-Higgs-model
recalling the structure of the background
gauge Lagrangian, the functional identities constraining the model
 and the normalization conditions for the amplitudes.
In section 3 we shall briefly discuss the physical functional variables.
In section 4  we shall define  the gauge invariant effective
action and  we shall discuss the
proof of the background equivalence theorem. The extent of our proof
is discussed in section 5.

\newsection{The reference model}

In this section we discuss the quantization rules of an $SU(2)$ Higgs
model \cite{BRS}. These rules, beyond the assignment of a classical action,
define the symmetry constraints, written in the form of functional
differential equations for the connected generator $\Z$ and the
effective action $\C$, and the normalization conditions for vertices
and propagators. To simplify the symmetry constraints we use the
Lautrup-Nakanishi auxiliary fields \cite{LN}  inserting them, as Lagrange
multipliers, in the gauge fixing term of the Lagrangian. Since the
Lagrangian is quadratic in these auxiliary fields, integrating over
them, leads directly to the conventional Feynman-'t Hooft gauge fixing.

Just to fix  symbols and functional variables we begin listing the
quantum fields and the background ones. For simplicity we avoid spinor
fields. Thus the theory is built in terms of the quantum fields:
\be
\varphi = {1\o \sqrt {2}}\(\begin{array}{c}
\pi_2 + i\pi_1\\
\sigma - i\pi_3\\
\end{array}\)\ , \quad \mbox{\boldmath $A$}_{\mu}\equiv (A^1_{\mu},
A^2_{\mu}, A^3_{\mu})\ .
\label{campi1}
\ee
The corresponding background fields are:
\be
\phi = {1\o \sqrt {2}}\(\begin{array}{c}
\Pi_2 + i\Pi_1\\
\Sigma - i\Pi_3\\
\end{array}\)\ , \quad \mbox{\boldmath $V$}_{\mu}\equiv (V^1_{\mu},
V^2_{\mu}, V^3_{\mu})\ .
\label{campi2}
\ee
Following Faddeev and Popov, the gauge fixing procedure requires the
introduction of a system of ghosts and anti-ghosts:
\be
\mbox{\boldmath $c$}\equiv (c^1, c^2, c^3)\ , \quad \bar{\mbox{\boldmath $c$}}
\equiv (\bar{c}^1, \bar{c}^2, \bar{c}^3)\ ,\label{campi3}\ee and of the above
mentioned Lautrup-Nakanishi multipliers:
\be
 \mbox{\boldmath $b$}\equiv
(b^1, b^2, b^3)\ .
\label{campi4}
\ee
The model is assumed to be invariant under background field gauge
transformations.
An infinitesimal background field gauge transformation is defined
by:
\bea
&&\delta_W\varphi = i{g\o 2}\mbox{\boldmath $\lambda$}\cdot \mbox{\boldmath
$\tau$}\(\varphi +
\tilde{v}\)\ ,\quad  \d_W\mbox{\boldmath $A$}_{\mu} = \na \mbox{\boldmath
$\lambda$}\ ,\nn
&&\delta_W\phi = i{g\o 2}\mbox{\boldmath $\lambda$}\cdot
\mbox{\boldmath $\tau$}\(\phi +
\tilde{v}\)\ , \quad \d_W\mbox{\boldmath $V$}_{\mu} = \nv \mbox{\boldmath
$\lambda$}\ ,
\label{W1}
\eea
where the nablas label covariant derivatives whose indices indicate
the corresponding connections and:
\be \tilde{v} = {1\o \sqrt {2}}\(\begin{array}{c}
0\\
v\\
\end{array}\)\ ,
\ee is the vacuum expectation value of the scalar field $\varphi$.
The second main symmetry property of our model  is BRS invariance.
In particular the classical action is assumed to be invariant under
the system of transformations:
\be
\d_S\mbox{\boldmath $A$}_{\mu} = \na \mbox{\boldmath $c$}\ ,\quad
\delta_S\varphi = i{g\o 2}\mbox{\boldmath $c$}\cdot \mbox{\boldmath
$\tau$}\(\varphi + \tilde{v}\)\ ,
\quad \delta_S\mbox{\boldmath $c$} = {g\o 2}\mbox{\boldmath
$c$}\w\mbox{\boldmath $c$}\ ,\quad
\delta_S\bar{\mbox{\boldmath $c$}} = \mbox{\boldmath $b$}\ , \quad
\delta_S\mbox{\boldmath $b$} = 0\ .
\label{ST1}
\ee
These transformations commute with the background gauge transformations.
As shown by
Grassi \cite{G1} it is convenient to extend to the background fields
the action of BRS transformations introducing a set of anticommuting
external fields:
\be
\Omega = {1\o \sqrt {2}}\(\begin{array}{c}
\Omega_2 + i\Omega_1\\
\Omega_4 - i\Omega_3\\
\end{array}\)\ ,  \quad \mbox{\boldmath $\Omega$}_{\mu}
\equiv (\U^1, \U^2, \U^3)\ ,
\label{campi5}
\ee
and defining:
\be
\delta_S\mbox{\boldmath $V$}_{\mu}= \mbox{\boldmath $\Omega$}_{\mu}\ , \quad
\delta_S\phi = \Omega\ ,  \quad
\delta_S\mbox{\boldmath $\Omega$}_{\mu} = 0\ , \quad \delta_S\Omega = 0\ .
\label{ST2}
\ee
 BRS transformations, being non-linear, transform the elementary
fields into composite operators; in the functional framework these
operators are coupled to external fields, that in the recent
literature are called anti-fields, and appear in the functional form
of the Slavnov-Taylor identity. In our case there are anti-fields
corresponding to $\mbox{\boldmath $A$}_{\mu}$\ , $\varphi$ and
$\mbox{\boldmath $c$}$\ , they
are:
\be
\mbox{\boldmath $A$}^{*}_{\mu}\equiv\(A^{*}_{\mu , 1}, A^{*}_{\mu , 2},
A^{*}_{\mu , 3}\)\ , \quad
\varphi^{*} \equiv {1\o \sqrt {2}} \(\begin{array}{c}
\varphi^{*}_{2} - i\varphi^{*}_{1}\\
\varphi^{*}_{4} + i\varphi^{*}_{3}\\
\end{array}\)\ , \quad e \quad \mbox{\boldmath $c$}^{*}\equiv\(c^{*}_{1},
c^{*}_{2}, c^{*}_{3}\)\ .
\label{campi7}
\ee
The action of the model is given by
\be
\C_0 = \int \({\cal{L}}_{inv} + {\cal{L}}_{g.f.} + {\cal{L}}_{\Phi.\Pi} +
{\cal{L}}_{S.T.}\)\ .\label{1}
\ee
The first term under integral is the well known gauge invariant Lagrangian
density of the $SU(2)$ Higgs model \cite{HK},
the second is the gauge fixing term:
\be
{\cal{L}}_{g.f.} = \mbox{\boldmath $b$}\nv\(\mbox{\boldmath
$A$}_{\mu}-\mbox{\boldmath $V$}_{\mu}\)
+ {\rho g\o 2}\mbox{\boldmath $b$}\[i\(\phi^{\dag} +
\tilde{v}\)\mbox{\boldmath $\tau$}
\(\varphi + \tilde{v}\) + h.c.\]+ {\mbox{\boldmath $b$}^2\o 2\alpha}\ ,
\label{1b}
\ee
and the third is the  Faddeev-Popov term; the last term
defines the BRS transformed fields through their
anti-field couplings:
\be
{\cal{L}}_{S.T.} = - \mbox{\boldmath $A$}^{* \mu}\na\mbox{\boldmath $c$}
-i\[\varphi^{*}{g\o 2}\mbox{\boldmath $\tau$}\(\varphi + \tilde{v}\) -
h.c.\]\mbox{\boldmath
$c$}
- \mbox{\boldmath $c$}^{*}{g\o 2}\mbox{\boldmath $c$}\w\mbox{\boldmath $c$}\ .
\ee
We shall not discuss here the technical aspects of the renormalization of
our model since this is a fairly well known
subject that the introduction of the background field does not change
substantially \cite{G1}. We assume therefore that the model
be quantized respecting all the symmetries of the classical action; this
implies, first of all, that the connected
functional satisfies the Slavnov-Taylor identity:
\be
{\cal{S}}\Z = \int\[\mbox{\boldmath $J$}_{\mu}{\d\o \d\mbox{\boldmath
$A$}^{*}_{\mu}}
+J_{\varphi_i}{\d\o \d\varphi^{*}_i}
+\bar{\mbox{\boldmath $\xi$}}{\d\o \d\mbox{\boldmath $c$}^{*}}
-\mbox{\boldmath $\xi$}{\d\o \d\mbox{\boldmath $J$}_b}
+\mbox{\boldmath $\Omega$}_{\mu}{\d\o \d\mbox{\boldmath $V$}_{\mu}}
+\Omega_i{\d\o \d\phi_i}\]\Z = 0\ ,
\label{S2}
\ee
where the functional variables $\mbox{\boldmath $J$}^{\mu}$, $J_{\varphi}$,
$\mbox{\boldmath $J$}_b$,
$\bar{\mbox{\boldmath $\xi$}}$ and $\mbox{\boldmath $\xi$}$ are the
classical sources
of the quantized fields $\mbox{\boldmath $A$}^{\mu}$\ , $\varphi$\ ,
$\mbox{\boldmath $b$}$\ ,
$\mbox{\boldmath $c$}$ and
$\bar{\mbox{\boldmath $c$}}$\ .

The invariance under the background gauge transformations (\ref{W1})
induces the Ward identity
\be \mbox{\boldmath $W$}\Z = 0\ , \label{ward}
\ee
where $\mbox{\boldmath $W$}$ is the differential operator generating
the transformations (\ref{W1}).
Since these gauge transformations commute with BRS ones,
 one has:\be
\[\mbox{\boldmath $W$}, {\cal{S}}\] = 0\ .
\label{com}
\ee
The gauge-fixing term (\ref{lt2}) is linear, in the sense that the
auxiliary field $\mbox{\boldmath $b$}$ multiplies an operator linear
in the quantized fields. In these conditions the auxiliary field equation
is a linear equation in the quantized
fields. Therefore this equation can be translated into a linear functional
differential equation for $\Z$, that
survives renormalization. In fact this equation is a renormalization
prescription for our model and is written
in the form:
\be
\mbox{\boldmath $J$}_b = \nv\({\d\Z\o\d \mbox{\boldmath $J$}_{\mu}}
-\mbox{\boldmath $V$}_{\mu}\)
+ {g\rho\o 2}\[i\(\phi^{\dag} + \tilde{v}\)\mbox{\boldmath
$\tau$}\({\d\Z\o \d J_{\varphi}}+
\tilde{v}\) + h.c.\] + {1\o \a}{\d\Z\o \d \mbox{\boldmath $J$}_{b}}\ .
\label{gf1a}
\ee
Combining this identity with the Slavnov-Taylor identity yields to a
further relation which is the BRS transformed of
(\ref{gf1a}):
\be\mbox{\boldmath $\xi$} = - \nv{\d\Z\o\d \mbox{\boldmath $A$}_{\mu}^{*}}
- \na\mbox{\boldmath $\Omega$}_{\mu}
- {g\rho\o 2}\[{i\o 2}\(\phi^{\dag} + \tilde{v}\)\mbox{\boldmath
$\tau$}{\d\Z\o \d\varphi^{*}}
+{i\o 2}\Omega\mbox{\boldmath $\tau$}\({\d\Z\o\d J_{\varphi}} + \tilde{v}\)
+ h.c.\]\ .
\label{sol2}
\ee
The equations (\ref{S2}), (\ref{ward}), (\ref{gf1a}) and (\ref{sol2}) give
a system of functional differential
constraints implementing the relevant symmetry properties of the fully quantized
version of our model. They can be translated
into a corresponding system of functional differential equations for the
effective action $\C$. In particular from
(\ref{gf1a}) and (\ref{sol2}) one has:
\be
{\d\C\o\d\mbox{\boldmath $b$}} = \nv\(\mbox{\boldmath $A$}_{\mu} -
\mbox{\boldmath $V$}_{\mu}\)
+ {\rho g\o 2}\[i\(\phi^{\dag} + \tilde{v}\)\mbox{\boldmath $\tau$}
\(\varphi + \tilde{v}\) + h.c.\] + {\mbox{\boldmath $b$}\o \alpha}\ ,
\label{gf1}
\ee
\be
{\d\C\o\d\bar{\mbox{\boldmath $c$}}}=\nv{\d\C\o\d\mbox{\boldmath $A$}^{*}_{\mu}}
-\na\mbox{\boldmath $\Omega$}_{\mu}
+ {\rho g\o 2}\[{i\o 2}\(\phi^{\dag} + \tilde{v}\)\mbox{\boldmath
$\tau$}{\d\C\o \d\varphi^{*}}
 + {i\o 2}\Omega\mbox{\boldmath $\tau$}\(\varphi + \tilde{v}\) + h.c.\]\ .
\label{gf2}
\ee
>From these equations one can extract some interesting information on $\C$.
Indeed the general solution to the system
(\ref{gf1}), (\ref{gf2}) has the following form:
\bea
\C &&=\bar{\C}\[\mbox{\boldmath $A$}_{\mu}, \mbox{\boldmath $V$}_{\mu},
\varphi, \mbox{\boldmath
$c$}, \tilde{\mbox{\boldmath $A$}^{*}_{\mu}}, \tilde{\varphi^{*}},
\mbox{\boldmath $c$}^{*}, \phi,
\mbox{\boldmath $\Omega$}_{\mu}, \Omega\]
\nn
&&+\int \[\mbox{\boldmath $b$}\nv\(\mbox{\boldmath
$A$}_{\mu}-\mbox{\boldmath $V$}_{\mu}\)
+ {\rho g\o 2}\mbox{\boldmath $b$}\[i\(\phi^{\dag} +
\tilde{v}\)\mbox{\boldmath $\tau$}
\(\varphi + \tilde{v}\) + h.c.\]+ {\mbox{\boldmath $b$}^2\o 2\alpha}\]\nn
&&-\int
\bar{\mbox{\boldmath $c$}}\[ \na\mbox{\boldmath $\Omega$}_{\mu} +i {\rho g\o 4}
\[\Omega\mbox{\boldmath $\tau$}\(\varphi +
\tilde{v}\) - h.c.\]\]
\ ,\label{sol1}\eea
where:
\be
\tilde{\mbox{\boldmath $A$}^{*}_{\mu}}=\mbox{\boldmath $A$}^{*}_{\mu}
- \nv\bar{\mbox{\boldmath $c$}}\ , \quad \tilde{\varphi}^{*}=\varphi^{*} -
i{g\rho\o 2}
\bar{\mbox{\boldmath $c$}}\(\phi^{\dag} + \tilde{v}\)\mbox{\boldmath
$\tau$}\ .
\ee
The functional $\bar{\C}$ is further constrained by the Slavnov-Taylor
identity that is:
\be
\int\[{\d{\bar{\C}}\o \d\mbox{\boldmath $A$}_{\mu}}
{\d{\bar{\C}}\o \d\tilde{\mbox{\boldmath $A$}^*}_{\mu}}+ {\d{\bar{\C}}\o
\d\varphi_i}
{\d{\bar{\C}}\o \d\tilde{\varphi^*_i}}+{\d{\bar{\C}}\o \d\mbox{\boldmath $c$}}
{\d{\bar{\C}}\o \d{\mbox{\boldmath $c$}^*}} + \mbox{\boldmath $\Omega$}_{\mu}
{\d{\bar{\C}}\o \d\mbox{\boldmath $V$}_{\mu}} + \Omega_i{\d{\bar{\C}}\o
\d\phi_i}\] = 0\ ,
\label{ST3}
\ee
and by the background Ward identity:
\bea
&&\mbox{\boldmath $W$}\bar{\C}\equiv \na{\d{\bar{\C}}\o \d\mbox{\boldmath
$A$}_{\mu}} +
\nv{\d{\bar{\C}}\o \d\mbox{\boldmath $V$}_{\mu}}-ig\(\(\varphi +
\tilde{v}\){\mbox{\boldmath $\tau$}\o
2}{\d{\bar{\C}}\o \d\varphi}-h.c.\) \nn
&&- ig\(\(\phi + \tilde{v}\){\mbox{\boldmath $\tau$}\o 2}
{\d{\bar{\C}}\o \d\phi}-h.c.\) + ...
=0\ .
\label{ward1}
\eea
The dots refer to the contribution of the anti-fields, and of
$\mbox{\boldmath $c$}$ ,
$\mbox{\boldmath $\Omega$}_\mu$ and $\Omega$.

 We can now discuss the parametrization
of our $\bar{\C}$. Considering the reparametrizations leaving
(\ref{ST3}) and (\ref{ward1})
unchanged, one has:
\bea
&&\mbox{\boldmath $A$}_\mu\rightarrow {1-k\o z_g}\mbox{\boldmath $A$}_\mu
+ {k\o z_g}\mbox{\boldmath $V$}_\mu\ ,
\quad \mbox{\boldmath $V$}_\mu\rightarrow{1\o z_g}\mbox{\boldmath
$V$}_\mu\ ,\quad \mbox{\boldmath
$\Omega$}_\mu\rightarrow {1\o z_g}\mbox{\boldmath $\Omega$}_\mu\ , \quad
\tilde{\mbox{\boldmath
$A$}^*}_\mu\rightarrow {z_g\o 1-k}\tilde{\mbox{\boldmath $A$}^*}_\mu\ ,\nn
&&\varphi\rightarrow
z_\varphi\(\varphi + {l\o 1-l}\phi\)\ , \quad  \phi\rightarrow{z_\varphi\o
1-l}\phi\ ,
\quad \Omega\rightarrow {z_\varphi\o 1-l}\Omega\ , \quad
\tilde{\varphi^*}\rightarrow
{1\o z_\varphi}\tilde{\varphi^*}\ , \nn
&&\mbox{\boldmath $c$}\rightarrow z_c\mbox{\boldmath $c$}\ , \quad
\mbox{\boldmath $c$}^*\rightarrow {1\o z_c}\mbox{\boldmath $c$}^*\ ,
\label{rip1}
\eea
accompanied by:
\bea
&&\bar{\C}\rightarrow \bar{\C} + \int \[k\ \mbox{\boldmath
$\Omega$}_\mu\tilde{\mbox{\boldmath
$A$}^*}_\mu + l\ \Omega\varphi\]\ ,\nn
&&g\rightarrow z_g g\ .
\label{rip2}
\eea
Since the theory is renormalizable it is easy to verify that, once the
parameters in (\ref{rip1})
have been fixed, one is left with a single free parameter corresponding to
the $\varphi^4$ coupling in
the invariant lagrangian. Therefore, taking also into account the parameters
appearing in the gauge-fixing,
one sees that the theory is identified by the dimensionless parameters
\be\a\ , \rho\ , g\ , \lambda\ , k\ ,
l\ , z_c\ , z_\varphi\ ,\label{par}\ee and by $v$\ . These parameters must
be fixed
by the
normalization conditions.

It is apparent from (\ref{rip1}) that neither $A_{\mu}$ nor $\varphi$ are
multiplicatively renormalized, while $Q_{\mu}\equiv A_{\mu} - V_{\mu}$
and $\varphi^q\equiv\varphi-\phi$ are. It is well known that $Q_{\mu}$
and $\varphi^q$ are the natural dynamical variables of the quantized theory
and, in fact, all the papers based on the background field method
choose the $Q$-framework, that is the natural dynamical variables
$Q_{\mu}$ and  $\varphi^q$
\cite{dW}.

However, as discussed in the introduction, the basic argument in
favour of the background gauge equivalence, that we consider in the
present paper, relies on the fact that, choosing the dynamical
variables $A_{\mu}$ and $\varphi$, the background fields become gauge
artifacts.

Then the question arises about the equivalence of the $Q$- and
$A$-frameworks.  To answer this question the first point to be clarified is
that, at least in perturbation theory, the effective action is a formal
power series in both quantum and background fields.  Indeed the basic
idea of the background method is to compute the amplitudes with only
background external legs, where the quantum fields ( $Q$ and $\varphi$
) contribute to the internal propagators.  These amplitudes are power
series in the background fields whose coefficients correspond to the
effective vertices.

In the functional framework Feynman amplitudes are obtained
renormalizing the Feynman vacuum functional whose functional integral
expression accounts for the diagrammatic structure of the amplitudes.
If $Z^{Q}$ is the vacuum functional in the $Q$-framework and $Z^{A}$
that in the $A$-framework, at the formal level of the unrenormalized
Feynman formula, it is clear that these functionals are related by: \be
Z^{A}\[ j^{\mu} , \ldots\]=e^{i\int \(j^{\mu}V_{\mu}+\ldots\)} Z^{Q}\[
j^{\mu} , \ldots\]\ .\label{tras}\ee since the ``bare'' actions of
both frameworks coincide.  It remains to verify what happens after
renormalization.

As discussed before, to renormalize our model corresponds to implement
the symmetry constraints that are written in the form: (\ref{S2}),
(\ref{ward}), (\ref{gf1a}) and (\ref{sol2}) in the $A$-framework and
can be easily translated into the form suitable for the $Q$-framework.  It
is also easy to verify that these constraints are compatible with
(\ref{tras}).  Once the symmetry constraints are implemented, two
different renormalization schemes differ in the parametrization; that
is, they correspond to different choices of the free parameters listed in
(\ref{par}).  This means that, given $Z^{Q}$, (\ref{tras}) defines a
$Z^{A}$ corresponding to a particular choice of the parameters in the
$A$-framework; in other words (\ref{tras}) defines a one-to-one
correspondence between the renormalized functional of each
framework, proving their equivalence.

 To complete the analysis of the $SU(2)$ Higgs-model we identify a
system of normalization conditions fixing the free parameters.
Assuming the notation:
${\d^{2}\C\o
\d
\Phi\d\Phi  `}\vert_{\Phi =0}\equiv\C_{\Phi\Phi'}$,
we assign the following wave function normalizations,
masses and couplings:
\bea
&&\C_{QQ}^{\mu,\nu}(q^2 = m_Q^2) = 0\ , \quad
\C_{\sigma^q\sigma^q}(q^2=M^2) = 0\ ,
\quad
\C_{\bar{c} c}(q^2=m_{\Phi\Pi}^2) = 0\ ,
\nn
&&\C_{QQ}^{\prime \mu,\nu}(q^2 = m_Q^2) =  g^{\mu\nu}\ , \quad
\C_{\sigma^q\sigma^q}^{\prime}(q^2=M^2) = 1\ , \quad
\C_{\bar{c} c}^{\prime}(q^2=m_{\Phi\Pi}^2) =  1\ ,
\nn
&&\C_{\sigma QQ}^{\mu\nu}(M^2, m_Q^2, m_Q^2) = g^{ph} m_Q g^{\mu\nu}\ , \quad
\C_{\Sigma\bar{c}c}(0, m_{\Phi\Pi}^2 \ , m_{\Phi\Pi}^2) = g^{ph}
{m_{\Phi\Pi}^2\o m_Q}\ .
\label{norm}
\eea
To avoid double poles in the propagators (\cite{BRS}, \cite{tH1}) we also
assume the
condition (`t Hooft):
\be
\left.\C^{\mu}_{Q_L b}\C_{b\pi}+\C_{bb}\C^{\mu}_{\pi Q_L}\right|_{q^2
=m_{\Phi\Pi}^2}=0 \ .
\label{tHo}
\ee
One can verify that the normalization conditions in (\ref{norm}) determine the
free parameters (\ref{par}) and $v$ ; in particular one has, up to one
loop corrections:
\bea
&&g=g^{ph}(1 + O(\hbar))\ ,\quad\lambda = {g^2M^2\o 2 m^2_Q}(1 +
O(\hbar))\ , \quad
v = {m_Q\o g}(1 + O(\hbar))\ , \nn
&& \rho = {2m_{\Phi\Pi}^2\o m_Q^2}(1 +
O(\hbar))\ , \quad
k = O(\hbar)\ , \quad  l =  O(\hbar)
\ ,\quad
z_{\varphi} = 1 + O(\hbar)\ , \quad z_c = 1 + O(\hbar)\ .\nonumber
\eea
One has also:
$$
{\rho\o 2} - {1\o \a} = O(\hbar)\ ,
$$
from which:
$$
\a = {m_Q^2\o m_{\Phi\Pi}^2}(1 + O(\hbar))\ .
$$
In Appendix A we list the propagators of our model.

\newsection{The physical variables}

Having specified the reference gauge model, we must discuss briefly its
physical content; that is the physical operators relevant for the
construction of the $S$-matrix.

First of all, the physical asymptotic fields  correspond
to the transverse components of the vector field and to the Higgs
field $\sigma$; then we must consider the composite
physical operators. We do not need a complete list of these operators;
we simply mention an example: the energy momentum tensor. We  associate
with every physical
operator a corresponding functional variable that is identified, in
the case of composite operators, with
the $\tau^{ph}$ external fields appearing in the introduction, and, in the
case of asymptotic fields, with  $\j$ in Eq.(\ref{jas}).

In a general gauge theory one defines a physical variable ($\Xi$) as a
functional variable with vanishing Faddeev-Popov charge, that is
coupled to a BRS-invariant operator  that does not correspond to
the BRS-variation of any other operator.
In formulae $\Xi$ is a physical variable if and only if:
\be \[\dX ,\S\] =0\
\label{p1}\ee
and
\be\dX\not= \{{\d\o\d X},\S\}\ .\label{p2}
\ee
Notice that the second condition (\ref{p2}) is crucial; indeed, for
example, the source of the auxiliary field $\mbox{\boldmath
$J$}_b$ and the background
field $\mbox{\boldmath $V$}_{\mu}$ satisfy (\ref{p1}), however one has:
$$
{\d\o \d\mbox{\boldmath $J$}_b} = \left\{{\d\o \d\mbox{\boldmath $\xi$}},
{\cal S}\right\}\ ,
\quad {\d\o \d\mbox{\boldmath $V$}_{\mu}}
= \left\{{\d\o \d\mbox{\boldmath $\Omega$}_{\mu}}, {\cal S}\right\}\ ,
$$
and hence these variables are physically trivial.

 Notice also that
 the actually independent asymptotic variables are the components of $\fp$
defined
in (\ref{jas}); we should therefore use $ {\cal Z}^{-1} K*\D_{+}*
f^{ph}$ instead of $\j$. For simplicity we prefer to use $\j$
understanding its dependence on $\fp$; however taking functional
derivatives we have to refer to $\fp$ using:
\be\df=  {\cal Z}^{-1} K*\D_{+}*\dj\ .\label{der}\ee

To simplify the notation in the rest of the paper we shall use the
following symbols:
\bea
&& j\equiv\(\mbox{\boldmath $J$}_{\mu}\ , J_{\varphi}\)\ , \quad
J\equiv\mbox{\boldmath $J$}_b\ , \quad \Omega\equiv\(\mbox{\boldmath
$\Omega$}_{\mu}\ , \Omega\)\ ,\nn
&& \Phi\equiv\(\mbox{\boldmath $A$}_{\mu}\ , \varphi\)\ , \quad
V\equiv\(\mbox{\boldmath $V$}_{\mu}\ , \phi\)\ , \quad
b\equiv \mbox{\boldmath $b$}\ ,\nn
&&\Phi^{*}\equiv \(\mbox{\boldmath $A$}^{*}_{\mu}, \varphi^{*}\)\ . \label{sim}
\eea
With these new symbols the Slavnov-Taylor identity
(\ref{S2}) becomes
\be
{\cal S}\Z \equiv\int \(j{\d\o \d \Phi^{*}} + \bar{\xi}{\d\o \d c^{*}} -
\xi{\d\o \d J}
+ \Omega{\d\o \d V}\)\Z = 0\ ,
\label{bb1}
\ee

\newsection{The effective action and the background equivalence theorem}

We begin  defining the effective action upon which
background gauge equivalence is based.

Since we are interested in the physical restriction of the $S$-matrix,
the ghost propagator does not appear.
Then we restrict our
functional variables setting: \be\Omega = \Phi^{*} = c^{*} = \bar{\xi} =
\xi = c=\bar{c}=0\ .\label{0}\ee
After this restriction the effective action, given in (\ref{sol1}), becomes:
\be \C\[\Phi,V,b\] =  \cq \[\Phi,V\] + \co\[ \Phi,b,V\]\ ,
\label{azione}\ee
where $\co$ contains the bosonic part of the gauge fixing term; in
the reference model:
\be\co =\int \[\mbox{\boldmath $b$}\nv\(\mbox{\boldmath
$A$}_{\mu}-\mbox{\boldmath $V$}_{\mu}\)
+ {\rho g\o 2}\mbox{\boldmath $b$}\[i\(\phi^{\dag} +
\tilde{v}\)\mbox{\boldmath $\tau$}
\(\varphi + \tilde{v}\) + h.c.\]+ {\mbox{\boldmath $b$}^2\o 2\alpha}\]\nn
\ .\label{gf}\ee
It is a crucial and general fact that the dependence of $\C$ on $b$ is
restricted to $ \co$.

To  simplify further our notation we shall  understand the dependence of the
connected functionals
and effective actions on the physical variables $\tau^{ph}$,
corresponding to physical composite operators, and we concentrate on the
asymptotic physical variables $\j$. Notice that these variables
appear  in the connected functional $\Z$ but not in $\C$.

Now we come to the main subject of this paper: the proof of background
equivalence following the lines presented in
the introduction.
As already discussed, we must compare the connected $S$-matrix
corresponding to the
effective action
(\ref{azione})  that is by no means invariant under gauge transformations
of $\Phi$ and $b$ at $V=0$ ,
with that corresponding to the alternative effective action:
\be
\C^{\prime}_{eff}\[\Phi\] =\cq\[\Phi, \Phi\] + \co\[\Phi, b , 0\]  \ ,
\label{eff}
\ee
which identifies our prescription for the gauge-fixed, gauge-invariant,
effective
action (\ref{ea}).
We  call this effective action {\it gauge invariant} since its gauge
invariance is only broken
by the gauge fixing term  ((\ref{gf}) at $V=0$) that is by the choice of
the propagators.
Indeed $\cq\[\Phi, \Phi\]$ is gauge invariant\footnote{But not gauge
independent! \cite{A}} owing to (\ref{ward1}).
The connected functional of our model is identified with the solution to:
\bea\Z\[j,J,V\]&&=-\cq\[\dj\Z\[j,J,V\], V\]\nn
&&-\co\[\dj\Z\[j,J,V\], \dJ\Z\[j,J,V\],V\]
+\int\(j\dj +J\dJ\)\Z\[j,J,V\]\ ,\label{t1}\eea
vanishing in the origin of the functional variable space;
while that corresponding to the gauge invariant effective action
(\ref{eff}) is identified with:
\bea\bZ\[j,J,V\]&&=-\cq\[\dj\bZ\[j,J,V\],
V+\dj\bZ\[j,J,V\]\]-\co\[\dj\bZ\[j,J,V\],
\dJ\bZ\[j,J,V\],0\] \nn &&+\int\(j\dj +J\dJ\)\bZ\[j,J,V\]\ ,\label{t2}\eea
The corresponding connected $S$-matrices are  given by $\Z\[\j,0,0\]$ and
$\bZ\[\j,0,0\]$ .

For background equivalence to hold true they should coincide.

To prove this, we introduce a further connected functional $\hZ$  depending
on two background fields
$V$ and $W$; $\hZ$ is defined by:
\bea\hZ\[j,J,V, W\]&&=-\cq\[\dj\hZ\[j,J,V,W\], V\]-
\co\[\dj\hZ\[j,J,V, W\], \dJ\hZ\[j,J,V, W\],W\]\nn &&
+\int\(j\dj +J\dJ\)\hZ\[j,J,V,W\]\ .\label{t3}\eea
We shall use $\hZ$ to verify the dependence of $\Z$  on the background
field appearing in $\co$.
It is obvious that:\be\hZ\[j,J,V, V\]=\Z\[j,J,V\] \ .\label{t4}\ee Taking the
$\Omega$-derivative of
(\ref{bb1}) in the point specified by (\ref{0})  and $j=\j$, we have:
\be\dV\Z\[\j , J, V\] =\(\dV +\dW\)\hZ\[\j,J,V, V\]=0\ ,\label{t5}\ee
since $\fp$ satisfies (\ref{p1}).
Taking the $\xi$ derivative in
the same conditions, we have:
\be\dJ\Z\[\j , J, V\] =\dJ\hZ\[\j,J,V, V\]=0\ .\label{t6}\ee
Using (\ref{ee}) we have:
\be\dW\hZ\[j,J,V,W\]=-\dW\co\[\dj\hZ\[j,J,V,W\],\dJ\hZ\[j,J,V,W\],W\]\
.\label{t7}\ee
The right-hand side of (\ref{t7}) can be easily computed taking into
account the explicit form of $\co$  given in (\ref{gf}). We exploit
in particular the fact that the  background
functional derivative of $\co$ is linear in the field $b$ :
\be \dV\co\[ \Phi, b,
V\]=L\[\Phi, V\] b\
,\label{t8}\ee
Combining (\ref{t7}) and (\ref{t8}), written as a functional differential
equation for $\hZ$, we
get:
\be \dW\hZ\[j,J,V,W\]= - L\[\dj\hZ\[j,J,V,W\], W\]\dJ\hZ\[j,J,V,W\]\
.\label{t9}\ee
Then starting from (\ref{t6}) and taking multiple $J$ and $W$
derivatives of (\ref{t9})
one finds recursively
that:
\be\(\dW\)^n \hZ\[\j,J,V,V\]=0\ ,\label{t10}\ee  for any n.
A more detailed analysis of this point is given in Appendix B.

>From (\ref{t5}) and (\ref{t6}) (see Appendix B), one finds that:\be
\hZ\[\j,J,V,W\]\equiv\Z\[\j,J,V\]
\equiv\Z\[\j,0,0\]\ .\label{t11}\ee
That is: the $S$-matrices corresponding to $\Z$ and $\hZ$ coincide.

Now we compare the  functional $\hZ$ with  $\bZ$. Setting $j=\j
$ and
$J=0$ and applying (\ref{lt2}) one finds:
\bea &&\(\dF
+\dV\)\cq\[\dj\bZ\[\j, 0,
V\], V+\dj\bZ\[\j, 0, V\]\]\nn &&+\dF\co\[\dj\bZ\[\j, 0, V\], \dJ\bZ\[\j,
0, V\], 0\]
=\j\ ,\nn &&\db\co\[\dj\bZ\[\j, 0, V\], \dJ\bZ\[\j,
0, V\], 0\]=0\
.\label{t12}\eea As mentioned in the introduction, this system determines
$\dj\bZ\[\j, 0, V\]$
and
$\dJ\bZ\[\j, 0, V\]$ uniquely.

One has furthermore from (\ref{ee}):
\be\dV\bZ\[\j, 0, V\]=-\dV\cq\[\dj\bZ\[\j, 0, V\], V+\dj\bZ\[\j, 0, V\]\]\
.\label{t13}\ee

In much the same way, considering $\hZ$ one has:
\bea &&\dF
\cq\[\dj\hZ\[\j, 0,
V,0\], V\]\nn &&+\dF\co\[\dj\hZ\[\j, 0, V,0\], \dJ\hZ\[\j, 0, V,0\], 0\]
=\j\ ,\nn &&\db\co\[\dj\hZ\[\j, 0, V,0\], \dJ\hZ\[\j, 0, V,0\], 0\]=0\
,\label{t14}\eea
that determine $\dj\hZ\[\j, 0, V,0\]$ and $\dJ\hZ\[\j, 0, V,0\]$ uniquely.

Furthermore from (\ref{t11}) and  (\ref{ee}) one has:
\be\dV\hZ\[\j, 0, V,0\]=-\dV\cq\[\dj\hZ\[\j, 0, V,0\], V\]=0\ .\label{t15}\ee

To compare $\hZ$ and $\bZ$ we consider the following system of functional
equations:
\bea \zeta\[\j , V\]&&=\dj\hZ\[\j, 0, V+\zeta\[\j , V\],0\]\ ,\nn
\eta\[\j , V\]&& =\dJ\hZ\[\j, 0, V+\zeta\[\j , V\],0\]\ .\label{t16}\eea
It is rather apparent that the iterative solution to (\ref{t16}) leads to a
unique solution ($\zeta$ ,
$\eta$). A  detailed analysis supporting this conclusion is given in Appendix B.

Therefore if we replace $V\rightarrow V+\zeta$ everywhere into the system
(\ref{t14}),
 on account of (\ref{t16}), we get:
\bea&&\dF\cq\[\zeta , V+\zeta \]+\dF\co\[\zeta , \eta , 0\]=\j\ ,\nn &&
\db\co\[\zeta , \eta , 0\]=0\ .\label{t17}\eea
Furthermore the same substitution into (\ref{t15}) gives:
\be \dV\cq\[\zeta , V+\zeta\]=0\ .\label{t18}\ee
Owing to  (\ref{t18}) we see that (\ref{t17}) is equivalent to :
\bea&&\(\dF+\dV\)\cq\[\zeta , V+\zeta \]+\dF\co\[\zeta , \eta , 0\]=\j\ ,\nn &&
\db\co\[\zeta , \eta , 0\]=0\ .\label{t19}\eea
Since this system  identifies uniquely its solution ($\zeta$ , $\eta$ ),
comparing (\ref{t19}) with
(\ref{t12}), we have:
\be
\zeta\[\j , V\]=\dj\bZ\[\j , 0, V\]=\dj\hZ\[\j, 0, V+\zeta\[\j , V\],0\]\
,\label{t20}\ee
\be\eta\[\j , V\]=\dJ\bZ\[\j , 0, V\]=\dJ\hZ\[\j, 0, V+\zeta\[\j , V\],0\]\
.\label{t21}\ee
If, using
(\ref{der}), we restrict the $j$-functional derivatives in (\ref{t20}) to the
physical shell
 and we take into account the $V$-independence of $\hZ\[\j,
0,V,0\]$ shown in (\ref{t11}), we get:
\bea\df\bZ\[\j , 0, V\]&&\equiv{\cal Z}^{-1} K*\D_{+}*\dj\bZ\[\j , 0,
V\]=\df\hZ\[\j, 0, V+\zeta\[\j , V\],0\]\nn &&=\df\hZ\[\j, 0, 0,0\]\
.\label{t22}\eea
Excluding the physical composite operators ($\tp=0$),  the last identity
can be integrated over
$\fp$ with the initial condition: $\hZ\[0,0,0,0\]=\bZ\[0,0,0\]=0$ ensuring, on
account of
(\ref{t11}), the identity of the connected $S$-matrices:\be \Z\[\j,0,0\]=
\hZ\[\j,0,0,0\]=
\bZ\[\j,0,0\]\ ,\label{t23}\ee
and  hence proving the background equivalence of the $S$-matrix
elements.
It is however possible to extend this results to the matrix elements
between physical
asymptotic states of T-ordered products of physical operators, proving that
$\hZ\[0,0,0,0\]=\bZ\[0,0,0\]$ for any choice of $\tp$.

This is easily done using (\ref{ee}), (\ref{t20}) and (\ref{t21}).
Indeed, applying (\ref{ee}) to $\bZ$ and $\hZ$, we get for $\j
=J=0$ :
\bea{\d \o \d \tau} \bZ\[0, 0, V\]&&= -{\d \o \d \tau}\cq\[\dj \bZ\[0, 0,
V\], V+\dj \bZ\[0, 0, V\]\]\nn &&-{\d \o \d \tau}\co\[\dj \bZ\[0, 0,
V\], \dJ\bZ\[0, 0, V\], 0\]\ ,\label{t24}\eea
and:
\bea{\d \o \d \tau} \hZ\[0, 0, V, 0\]&&= -{\d \o \d \tau}\cq\[\dj \hZ\[0, 0,
V,0\], V\]\nn &&-{\d \o \d \tau}\co\[\dj \hZ\[0, 0,
V,0\], \dJ\hZ\[0, 0, V,0\], 0\]\ .\label{t25}\eea
If we replace in (\ref{t25}) : $V\rightarrow V+\dj \bZ\[0, 0, V\]$, the
left-hand
side  does not change, owing to (\ref{t15}), and, on
account of (\ref{t20}) and (\ref{t21}),  the
right-hand side becomes equal to that of  (\ref{t24}). We thus
conclude that:
\be{\d \o \d \tau} \bZ\[0, 0, V\]={\d \o \d \tau} \hZ\[0, 0, V, 0\]\
.\ee Integrating with respect to $\tau$ with the initial condition:
$\hZ\[0,0,0,0\]=\bZ\[0,0,0\]=0$, we prove the identity: $\hZ\[\j,0,0,0\]=
\bZ\[\j,0,0\]$, and hence (\ref{t23}) for any $\tp$.

\newsection{Comments}

We stress, first of all, that our proof is based on the existence of a
fully renormalized theory satisfying a  set of renormalization
prescriptions ((\ref{S2}), (\ref{ward}), (\ref{gf1a}) and
(\ref{sol2})); the only explicit references to the perturbative
construction concern the reference model and the discussion of the
existence and uniqueness of the solution of the system (\ref{t16}).

The use of the simplified symbols introduced in (\ref{sim}) should
 put into evidence the general nature of our proof.  Indeed the
essential ingredients of the analysis can be divided into two sets: the
basic, physical, ingredient is Slavnov-Taylor identity (\ref{bb1})
ensuring that the background field and the auxiliary field are gauge
artifacts and have no influence on the physical amplitudes.  The
second, technical, ingredient is the linear gauge choice, that has
allowed us to separate from the effective action the gauge fixing part
(see (\ref{1}), (\ref{1b}) and (\ref{sol1})) guaranteing in particular
the property (\ref{t8}).  Of course everything is based on the
systematic use of the functional framework and in particular on
(\ref{lt2}) and (\ref{ee}) whose validity is completely general.

Therefore it should be clear that our proof extends directly to any
gauge model, provided one can define the asymptotic fields and hence
the $S$-matrix.  In fact, in the models in which the gauge group
contains abelian invariant factors there are further constraints that
are conveniently  introduced to guarantee the radiative stability of
abelian charges (\cite{BBBC} and \cite{G2}). These constraints, that
correspond to the prescription of the field equations of the abelian
anti-ghosts, further specify the gauge fixing prescription without any
interference with the ingredients of our proof.

A further point that requires a short discussion concerns the
dependence of the gauge invariant effective theory on the gauge fixing
prescription.  First of all, we should notice that our construction is
based on two, in principle independent, gauge fixing procedures.  The
first  {\it quantum} gauge fixing is needed to compute from the
lagrangian the effective action, the second one allows the construction
of the $S$-matrix from the effective action\footnote{the existence
of two distinct gauge fixings justifies the introduction of the
{\it interpolating} functional $\hZ$ depending on two background
fields $V$ and $W$.}.  It has been convenient
for us to identify these gauge fixings, since we had to compare the
$S$-matrix of the effective theory to that obtained directly from the
theory in a trivial background.  However it is  fairly well known
 that, once a gauge invariant effective action is given, the
$S$-matrix is independent of the gauge fixing necessary to define the
effective theory propagators.  It is also independent of the first,
quantum, gauge fixing, since this is true for the theory in a trivial
background \cite{BRS}, however, it is known that the gauge invariant effective
action is not \cite{A}.  This could appear a little paradoxical, since one
could
think that all the parameters appearing into a gauge invariant
effective action should carry an independent physical information, but
it is easy to show that this is not true, and there is wide room to
change the gauge invariant effective couplings that are proportional to the
field equations, thus keeping the
$S$-matrix fixed. Concerning the parametrization of gauge effective
actions see also \cite{GW}.

\eject
\noindent
{\bf Acknowledgements}

C.B. is deeply indebted to G.Curci for attracting his attention on the
phenomenological relevance of background equivalence; C.B. is also
indebted to R.Stora for a critical reading of a very preliminary
version of the manuscript.

This work is partially supported by the European Commission, TMR
programme ERBFMRX -
CT960045 and by MURST.

\renewcommand{\thesection}{A}
\renewcommand{\thesubsection}{A.\arabic{subsection}}

\vspace{12mm}
\pagebreak[3]
\setcounter{section}{1}
\setcounter{equation}{0}
\setcounter{subsection}{0}
\setcounter{footnote}{0}

\begin{flushleft}
{\bf Appendix A}
\end{flushleft}

\noindent
Taking into account the normalization conditions (\ref{norm}),
 the  propagators of the reference model are:
\bea
&&G^{QQ}_{\mu\nu}(q)
= {I\o q^2 - m_Q^2}\(g_{\mu\nu} - \({m^2_Q + m^2_{\Phi\Pi}\o m^2_Q}\)
{q_{\mu}q_{\nu}\o q^2-m^2_{\Phi\Pi}}\)\ \ ,
\nn
&&G^{\pi^q Q_L}_{\nu}(q) = 0\ , \quad
G^{b Q_L}_{\nu}(q) =  i{q_{\nu}\o q^2-m^2_{\Phi\Pi}}\ ,
\nn
&&G^{\pi^q b}(q)
= G^{b \pi^q}(q) = {m_Q\o q^2-m^2_{\Phi\Pi}}\ ,\quad
G^{b b}(q) = 0\ ,
\nn
&&G^{\pi^q\pi^q}(q)
= {1\o q^2-m^2_{\Phi\Pi}}\ , \quad G^{\sigma^q\sigma^q} = {1\o q^2-M^2}\ ,
\nn
&&G^{\bar{c}c}(q) ={I\o q^2-m^2_{\Phi\Pi}}\ .
\label{pp1}
\eea

\renewcommand{\thesection}{B}
\renewcommand{\thesubsection}{B.\arabic{subsection}}

\vspace{12mm}
\pagebreak[3]
\setcounter{section}{1}
\setcounter{equation}{0}
\setcounter{subsection}{0}
\setcounter{footnote}{0}

\begin{flushleft}
{\bf Appendix B}
\end{flushleft}

\noindent

We begin this appendix considering (\ref{t5}) , (\ref{t6}) and
(\ref{t9})
and proving recursively (\ref{t10}) for any n.

First of all, from (\ref{t6}) and (\ref{t9}) we have:
\be\dW\hZ\[\j , J, V ,V\]=0\ .\label{ab1}\ee
Let us assume (\ref{t10}) to hold true for any $n\leq m-1$,
up to order $m-1$ we have:
\be\dJ\(\dW\)^{n}\hZ\[\j , J, V ,V\]=0\ .\label{ab2}\ee
We can compute the $m^{th}$ $W$-derivative of $\hZ\[\j , J, V ,W\] $
for $V=W$ taking the $(m-1)^{th}$
$W$-derivative of both sides of (\ref{t9}) and putting $W=V$.
The right-hand side of the resulting equation is the sum of many terms,
each proportional to  a derivative
(\ref{ab2}) with $n< m$, therefore it vanishes and hence (\ref{t10}) holds true
for any $n$.

We notice furthermore that $\hZ\[\j , J, V,W\]$ is independent of $V$
and $W$. Indeed, Taylor expanding this functional around $V=W$, and
using  (\ref{t10}), we see that it is independent of $W$.
Then, on account of (\ref{t4}), it coincides with $\Z\[\j , J, V\]$ which,
according to (\ref{t5}), is $V$-independent. Thus we have proved
(\ref{t11})

Now we consider the system
 (\ref{t16}).
 For our purposes it is sufficient to study the iterative
solutions of this system that are formal power series in
$V$ and $\j$,
since the physical amplitudes, that we are considering, are identified
with
the coefficients of an analogous series. The iterative solution of
(\ref{t16})
is built developing the right-hand side of this equation
in series of $\zeta$ getting:
\bea
&&\zeta\[\j, V\](x)={\d\o\d J(x)}\hZ \[\j, 0, V,0\]\nn
&&+\int dy {\d ^2\o\d J(x)\d V(y)}\hZ \[\j, 0,V,0\] \zeta[\j, V](y) +
O\(\zeta ^2\) \ ,
\label{ab3}
\eea
that can be written in the form:
\bea
&&\int dy \[\d (x-y)-{\d ^2\o\d J(x)\d V(y)}
\hZ \[\j, 0,V,0\]\]\zeta[\j, V](y)\nn
&&={\d\o\d J(x)}\hZ \[\j, 0,V,0\]+O\(\zeta ^2\)\ .
\label{ab4}
\eea
Now it is clear that (\ref{t16}) is solvable provided the "matrix"
\be
\d (x-y)-{\d^2\o\d J(x)\d V(y)}\hZ\[\j, 0,V,0\] \ ,
\label{ab5}
\ee
 be non-degenerate at  $V = 0$ .
This is certainly true in perturbation theory,
 since the second term in the left-hand side of (\ref{ab4})
vanishes
in the tree approximation. Indeed, owing to (\ref{ee})
 and (\ref{t3}) , $\dV\hZ$ can be computed from $\dV\cq$ and $\cq$
 in the tree approximation reduces to $ \int {\cal{L}}_{inv}$, the
classical action
deprived of the gauge fixing part, that is independent of $V$.

 It is also clear that,
the $\zeta$-component of the solution of (\ref{t16})
 identifies uniquely the $\eta$-component.  Thus, at least
perturbatively, the system
(\ref{t16}) has a unique solution.

\vfill\eject

\end{document}